\documentstyle[fps,twoside,fleqn,espcrc2]{article}
\newcommand{\un}{1\!\!1}
\newcommand{\half}{\frac{1}{2}}

\newcommand{\lag}{\langle}
\newcommand{\rag}{\rangle}
\newcommand{\gm}{\gamma}

\newcommand{\Dm}{D_{\mu}^+}
\newcommand{\Dbm}{D_{\mu}^-}
\newcommand{\Pm}{\partial_{\mu}^+}
\newcommand{\Pbm}{\partial_{\mu}^-}

\newcommand{\ep}{\epsilon}

\newcommand{\kp}{\kappa}

\newcommand{\Ps}{\Psi}
\newcommand{\Ph}{\Phi}

\newcommand{\Om}{\Omega}
\newcommand{\Psb}{\overline{\Ps}}

\newcommand{\dsl}{\partial \!\!\!/}

\newcommand{\mn}{m_F^{(n)}}
\newcommand{\mc}{m_F^{(c)}}
\newcommand{\mdn}{m_D^{(n)}}
\newcommand{\mdc}{m_D^{(c)}}
\newcommand{\En}{E_F^{(n)}}
\newcommand{\Ec}{E_F^{(c)}}
\newcommand{\Edn}{E_D^{(n)}}
\newcommand{\Edc}{E_D^{(c)}}

\newcommand{\Pc}{\Psi^{(c)}}
\newcommand{\Pn}{\Psi^{(n)}}
\newcommand{\psn}{\psi^{(n)}}

\newcommand{\psbn}{\overline{\psi}^{(n)}}

\newcommand{\Pnb}{\overline{\Psi}^{(n)}}
\newcommand{\Pcb}{\overline{\Psi}^{(c)}}

\newcommand{\hmu}{\hat{\mu}}

\newcommand{\ra}{\rightarrow}

\newcommand{\be}{\begin{equation}}
\newcommand{\ee}{\end{equation}}
\newcommand{\bea}{\begin{eqnarray}}
\newcommand{\eea}{\end{eqnarray}}

\newcommand{\eq}{\ref}
\newcommand{\beq}{\begin{equation}}
\newcommand{\eeq}{\end{equation}}
\newcommand{\cc}{\cite}
\newcommand{\lb}{\label}

\def \3{\ss}

\newcommand{\AmS}{{\protect\the\textfont2
  A\kern-.1667em\lower.5ex\hbox{M}\kern-.125emS}}

\title{Chiral fermions in two dimensions?%
\thanks{Presented by Asit K. De.} }

\author{Wolfgang Bock
\address{Institute of Theoretical Physics, University of
         Amsterdam, Valckenierstraat 65, 1018 XE Amsterdam,
         The Netherlands},
        Asit K. De
\address{Washington University, Department of Physics,
        St. Louis MO 63130, USA},
         Erich Focht
\address{HLRZ c/o KFA J\"ulich, P.O. Box 1913, 5170 J\"ulich,
         Germany and \\
         Institute of Theoretical Physics E, RWTH
         Aachen, Sommerfeldstr., 5100 Aachen, Germany} and
         Jan Smit$^{\;\;{\rm a}}$
         }
\begin{document}

\begin{abstract}
Quenched studies of a global U(1) symmetric Wilson-Yukawa model in two
dimensions show no evidence of a charged fermion
in the vortex phase at strong
Wilson-Yukawa coupling while there is strong
indication of a massive neutral
fermion. However,
with the U(1)$_L$ gauge field turned on,
we use dimensional arguments to suggest that the neutral fermion appears
to couple chirally to a massive vector boson state.
\end{abstract}
%
\maketitle
\section{INTRODUCTION}
The Wilson-Yukawa (W-Y) approach to chiral gauge theories on the
lattice uses a gauge-invariant Wilson term, called the W-Y
coupling term, which contains radially frozen scalar
fields. The proposal fails for weak
W-Y coupling $w$;
only for strong $w$ the doubler fermions can
be decoupled completely by making them heavier than the cut-off. Based on
the results of our numerical studies and comparison with available
analytic results some of us concluded that the
approach still fails because the symmetric phase of the theory
at strong $w$ contains
only a massive neutral fermion whose couplings vanish as $a^2$.

The present work \cite{2d} investigates the same approach in two dimensions.
We find similar results as in four dimensions \cite{4d},
namely that there does not
seem to exist a charged fermion and there are very strong indications
for a massive neutral fermion.
In two dimensions, however, there is an interesting possibility suggested
from naive power counting that the effective interaction of
the neutral fermion might survive the continuum limit in the form of a
chiral coupling to a vector boson state. We explore this issue and find
positive evidence for an effective chiral coupling in two dimensions,
although the theory
emerges out to be very different from the original target addressed
in the W-Y approach.
\section{THE MODEL AND ITS PHASE DIAGRAM}
The action of a W-Y model in two euclidean dimensions may be written as:
\bea
S\!\!\!\!\!&=&\!\!\!\! \sum_x {\cal L} \;,\;\; \;
            {\cal L} = {\cal L}_{U}
                     + {\cal L}_{\Phi}
                     + {\cal L}_{\Psi},     \lb{S}  \\
%
{\cal L}_{\Psi} \!\!\!\!\!&=& \!\!\!\!\!\frac{1}{2} \sum_{\mu=1}^{2}
        \Psb \gm_{\mu} [(\Dm+\Dbm)P_L + (\Pm+\Pbm)P_R ] \Ps \nonumber \\
 &+&y \Psb (\Phi P_R + \Phi^* P_L) \Ps  \nonumber \\
                      &-& \frac{w}{2} (\Psb \Phi)
P_R \sum_{\mu=1}^2 \Pm \Pbm \Ps \nonumber \\
&-& \frac{w}{2} \Psb P_L \sum_{\mu=1}^2 \Pm \Pbm (\Phi^* \Ps)
\;. \lb{LFERM}
\eea
 ${\cal L}_{U}$ gives the usual plaquette action with gauge
coupling $g$ and
 ${\cal L}_{\Phi}$ is the usual lattice lagrangian
for radially frozen scalar fields $\Phi_x$ with
$\kappa$ as the scalar hopping parameter. ${\cal L}_{\Psi}$ includes the
W-Y coupling $w$ and an usual Yukawa coupling $y$.
Furthermore $\Dm  \Ps_x = U_{\mu x} \Ps_{x+\hmu} -\Ps_x$,
$\Dbm \Ps_x = \Ps_x - U_{\mu x-\hmu}^* \Ps_{x-\hmu}$,
$\Pm=\Dm|_{U=1}$ and $\Pbm=\Dbm|_{U=1}$.
$P_{R,L}=\frac{1}{2} (\un \pm \gm_5)$ with
$\gm_5=-i \gm_1 \gm_2$.

The action (\eq{S}) is invariant under the local gauge
transformations: $\Ps_{L,x} \ra \Omega_{L,x} \Ps_{L,x}$,
$\Psb_{L,x} \ra \Psb_{L,x} \Omega_{L,x}^*$,
$\Phi_x \ra \Om_{L,x} \Phi_x$, $U_{\mu x} \ra \Om_{L,x} U_{\mu x} \Om_{L,
x+\hmu}^*$, with $\Om_{L,x} \in$U(1)$_L$. It is furthermore
invariant under the global transformations:
$\Ps_{R,x} \ra \Omega_{R} \Ps_{R,x}$,
$\Psb_{R,x} \ra \Psb_{R,x} \Omega_{R}^*$ and
$\Phi_x \ra \Phi_x \Om_{R}^*$ with $\Om_{R} \in$U(1)$_R$.

In our numerical simulations, we have used the quenched approximation,
and all the calculations involving fermions are performed in the global
limit $g=0$ and $U_{\mu x}=1$. In this case the scalar part of the theory
reduces to the XY model.

The XY model is known to have a
phase transition at $\kp=\kp_c \approx 0.56$ which
separates a
vortex (VX) phase ($\kp < \kp_c$) with finite scalar correlation length
from a spin-wave (SW) phase where the scalar
correlation length is infinite. In practice, the SW and the VX
phase behave on a finite lattice exactly as the broken (FM) and the
symmetric (PM) phase of the four dimensional model.

In the quenched approximation $\kp_c$ is independent of $y$ and $w$.
However, as in four dimensions,
the fermionic sector has a crossover at
$y+dw\approx \sqrt{d/2}$, i.e. in $d=2$ at $y+2w\approx 1$ separating
weak coupling phases SW(W) and VX(W) from the strong coupling phases
SW(S) and VX(S). The behavior of the fermion mass changes across the
crossover and in the strong coupling phases, contrary to `expectations',
the fermion mass increases as $\kp$ decreases. Only in the strong coupling
phases the doublers can be decoupled by making them heavier than the cutoff
and we perform numerical simulations particularly in the VX(S) phase.
(The VX(W) phase does not seem to be
present in the full model with dynamical fermions and could be
an artefact of the quenched
approximation \cc{DeFo92}. However, this does not affect our
investigations in the VX(S) phase.)
\section{FERMION MASSES AND HOPPING PARAMETER EXPANSION}
In two dimensions, according to the well-known
Mermin-Wagner-Coleman theorem,
there cannot be a spontaneous breakdown of
a continuous symmetry, the U(1)$_L\otimes$U(1)$_R$ chiral symmetry
in our case. However, two Dirac fields $\Pn$ and $\Pc$ may be constructed
which transform vectorially under U(1)$_R$ and U(1)$_L$ respectively:
\be
\Pn = ( \Phi^{*}P_L + P_R ) \Ps,\;
\Pnb =\Psb (\Phi P_R + P_L) ,
\lb{NEU}
\ee
\be
\Pc = (P_L + \Phi P_R) \Ps ,\;
\Pcb=\Psb (P_R + \Phi^{*} P_L) .
\lb{CHA}
\ee
$\Pn$ and $\Pc$ will henceforth be called the neutral and the charged fermion
field respectively because of their behavior under the U(1)$_L$ , the group
gauged in the action (\eq{S}). Since these fields transform vectorially,
mass terms can be written down for them.

Calculation of neutral and charged fermion propagators form the bulk
of this work. First we present results for the propagators from the
fermionic hopping parameter expansion (HPE). To effect the HPE, the fermionic
Lagrangian (\eq{LFERM}) is rewritten in terms of $\Pn$:
\bea
{\cal L}_F \!\!\!&=& \!\!\! \half \sum_{\mu=1}^2 [
(\Pnb_L \Phi^*) \gm_{\mu}(\Dm+\Dbm) (\Phi \Pn_L) \nonumber \\
&+& \Pnb_R \gm_{\mu}(\Pm+\Pbm) \Pn_R ] \nonumber \\
&+& y \Pnb \Pn - \frac{w}{2} \Pnb \sum_{\mu=1}^2 \Pm \Pbm \Pn \;.\lb{LN}
\eea

Using the lagrangian (\eq{LN}) one finds,
to lowest order in the hopping parameter $\alpha = 1/(y+2w)$,
an expression for the neutral fermion propagator in
momentum space, from which one can read off the fermion masses
$\mn$ and $\mdn$ for the physical fermion and the species doublers:
\be
\mn  \approx  y z^{-1},\;
\mdn \approx \mn + 2 w l z^{-1} ,\; l=1,2 ,
\lb{MN}
\ee
where $l$ is the number of momentum components equal to $\pi$ in the
two dimensional Brillouin zone. The quantity
$z^2  = \lag \mbox{Re} (\Ph_x^{*} U_{\mu x} \Ph_{x+\hmu}) \rag$ has
a non-vanishing value in both VX and SW phases.

Writing the action first in terms of the charged fermion and then doing
HPE for the charged fermion propagator,
similar formulas can be obtained
for the masses of the charged fermion (assuming it exists for the moment)
and its species doublers
\be
\mc  \approx (y+4w) z^{-1} -4wz ,\;
\mdc \approx \mc + 2 w z l .
\lb{MC}
\ee
\section{COMPARISON OF THE NUMERICAL RESULTS WITH THE HPE}
We compute numerically the neutral and the charged
fermion propagators on equilibrated scalar field configurations and
analyze them in terms of the free Wilson fermion ansatz to extract
the rest energies $\En$ and $\Ec$
of the neutral and the charged fermion and the corresponding
rest energies $\Edn$ and $\Edc$ for the lowest lying doubler fermions.
In fig.~1 we have displayed the numerical values
for $\En$, $\Edn$, $\Ec$ and $\Edc$ as a function of $y$
for $\kp=0.4$ and $w=2.0$, which is well inside the VXS phase. The dashed,
solid, dash-dotted and dotted lines
correspond respectively to the HPE results for the
rest energies $\En$, $\Edn$, $\Ec$ and $\Edc$, as obtained from formulas
(\eq{MN}) and (\eq{MC}) and then using lattices dispersion relations.
The figure shows that the agreement between the numerical result and
the analytic prediction is quite
impressive for the rest energies $\En$ and $\Edn$
while the HPE curves for $\Ec$ and $\Edc$ exhibit a strong
deviation from the numerical results. In the case of $\Ec$ the deviation
is larger than a factor two. The figure shows furthermore that
$\En$ appears to vanish in the limit $y \ra 0$, in agreement with
the GP shift symmetry mentioned before,
whereas $\Edn$ stays above 1 for all values
of $y$ which implies the decoupling of the species doublers
of the neutral fermion in the continuum limit.
%
%
%
\begin{figure}[t]
\centerline{
\fpsxsize=6.5cm
\fpsbox{Taaa.ps}
}
\vspace*{-1.6cm}
\caption{ \noindent {The rest energies $\En$, $\Edn$, $\Ec$ and
$\Edc$ as a function of $y$ for $\kp=0.4$ and $w=2.0$.}}
\label{aaa}
\end{figure}
We refer the reader to ref. \cite{2d} for a full account of the numerical
results.

\section{EFFECTIVE THEORY AT STRONG W-Y COUPLING $w$}
The excellent agreement of the numerical data with the HPE predictions
for the neutral fermions suggests that
the physics in the strong coupling region is well described
by the lagrangian (\eq{LN}) in terms of the neutral fermion fields.
The charged fermion fields $\Pc=\Ph\Pn$ and
$\Pcb=\Pnb \Phi^*$ can then be regarded as composite fields and
the charged fermion, provided it exists at all in the particle
spectrum, is to be considered as a bound state
with binding energy $\ep_B$ defined as
\be
\Ec=\En+E_{\Phi}+\ep_B     \;,
\ee
where $E_{\Phi}$ is the rest energy of the scalar.
\begin{table}[hbt]
\setlength{\tabcolsep}{0.45pc}
\newlength{\digitwidth} \settowidth{\digitwidth}{\rm 0}
\catcode`?=\active \def?{\kern\digitwidth}
\caption{Various rest energies and $\ep_B$ in dependence of $L$
         on lattices $L\times 64$ at the point
$(\kp,y,w)=(0.45,0.3,2.0)$. The errors of $\En$ are negligible.}
\label{tab:TAB1}
\begin{tabular}{ccccc}
\hline
$L$  &  $E_{\Phi}$ & $\En$   & $\Ec$        & $\ep_B$     \\
\hline
$16$ &  $0.126(3)$ & $0.275$ & $0.368(13)$ & $-0.033(16)$ \\
$32$ &  $0.115(3)$ & $0.277$ & $0.374(9) $ & $-0.018(12)$ \\
$48$ &  $0.120(5)$ & $0.276$ & $0.379(12)$ & $-0.017(17)$ \\
$64$ &  $0.119(5)$ & $0.277$ & $0.389(11)$ & $-0.007(16)$ \\
\hline
\end{tabular}
\end{table}

 Table 1 shows the numerically obtained
values of the various rest energies and the binding energy at a point in
parameter space inside the VX(S) phase for various $L$ on $L\times 64$
lattices. The binding energy is
clearly very small and  almost
compatible with zero even on the smaller lattices. Furthermore there is
a systematic trend of $|\ep_B|$ to
decrease when $L$ increases. This along with many other
similar pieces of evidence
in ref. \cite{2d} indicate that
the signal detected in the charged fermion propagators can very well be
just a two particle state of the neutral fermion
and the scalar particle.

For $g>0$ the VX phase is expected to turn into
a confinement phase where scalar particles
confine into massive bosonic particles. Writing the effective gauge field
combination $U_{\mu x}^{\prime}=
\Phi^*_x U_{\mu x} \Phi_{x+\hmu}$ in (\eq{LN}) in the standard fashion:
$U_{\mu x}^{\prime}=z^2 +  H_{\mu x}  + i W_{\mu x}
\rightarrow  z^2 +  m_H  H_x  + i W_{\mu x}$,
where $H_x$ and  $W_{\mu x}$ are interpolating fields for the scalar and
vector bosonic bound states in the confinement phase and
 $m_H$ is some mass scale (introduced for dimensional reason) and after
 a trivial rescaling of the fields $\Pn_L$ and $\Pnb_L$
we obtain for ${\cal L}_F$ the form
\bea
&{\cal L}_F& = \half \sum_{\mu=1}^2 \left[
               \Pnb_x  \gm_{\mu}  \Pn_{x+\hmu} -
               \Pnb_{x+\hmu}  \gm_{\mu}  \Pn_{x} \right] \nonumber \\
&+&\!\!\!\!\! \frac{y}{z} \Pnb \Pn
        - \frac{w}{2z} \Pnb \sum_{\mu=1}^2 \Pm \Pbm \Pn
      \lb{e1} \\
&+&\!\!\!\!\! \frac{m_H}{z^2} \half \sum_{\mu=1}^2 H_x [
      \Pnb_{L,x}  \gm_{\mu} \Pn_{L,x+\hmu}\nonumber \\
&-& \Pnb_{L,x+\hmu}  \gm_{\mu} \Pn_{L,x} ] \lb{e2} \\
&+&\!\!\! \!\!\frac{1}{z^2} \half \sum_{\mu=1}^2 i W_{\mu x} [
      \Pnb_{L,x}  \gm_{\mu} \Pn_{L,x+\hmu} \nonumber \\
&+&   \Pnb_{L,x+\hmu}  \gm_{\mu} \Pn_{L,x} ]\lb{e3}
\eea
In the above (\eq{e1})
describes a free neutral fermion with mass $\mn =y/z$, (\eq{e2})
 suggests that the coupling of the neutral fermion
to the Higgs-like bound state vanishes like $a$. However, in (\eq{e3})
the neutral fermion couples chirally to the
vector boson field $W_{\mu x}$ if its
dimension is one, as suggested by the naive dimensional analysis.

In order to find out whether the fields $W_{\mu x}$ and $H_{\mu x}$
are indeed dimension one operators we have computed numerically the
scale dependence of the corresponding wave-function renormalization constants
$Z_H$ and $Z_W$ in the gauge-Higgs system.
Fig.~2 collects our results and displays strong support for the naive
dimensionalities.

%
%
\begin{figure}[t]
\centerline{
\fpsxsize=6.0cm
\fpsbox{Tbbb.ps}
}
\vspace*{-1.3cm}
\caption{ \noindent {
The wave-function renormalization constants $Z_W$ (squares)
and $Z_H$ (circles) are plotted respectively as a function of
$m^2=a^2 m_{W,phys}^2$ and $m^2=a^2 m_{H,phys}^2$
for the fixed ratio $m_H/m_W=1.14$.
The dotted lines are to guide
the eye.}}
\label{bbb}
\end{figure}
The fermion couplings in the VXS phase may then be summarized
qualitatively by the following effective lagrangian
\bea
{\cal L}_F^{eff} &=& \psbn \dsl \psn +\mn \psbn \psn  \nonumber \\
&+& i g_R \psbn_L \gm_{\mu} \psn_L  W_{\mu}^{(c)} \;, \lb{EFF2}
\eea
where $g_R=\sqrt{Z_{W}}/z^2$, $\psn_L$ etc. are continuum fermionic
fields and
$W_{\mu}^{(c)}$ is the
vector field in the continuum with standard normalization.

Although the original target of the W-Y approach is not achieved,
we end by noting that in two dimensions the neutral fermion exhibits
indeed a non-vanishing chiral coupling to the massive vector boson; however,
in the global limit there does not seem to be a strong enough $\Pn$-$\Phi$
interaction to produce a charged fermion. \\

We thank M.F.L.~Golterman, J.~Jers\'ak
and D.N.~Petcher for valuable discussions.
The simulations           were performed
on the CRAY Y-MP/832 at HLRZ J\"ulich and on the CRAY Y-MP4/464
at SARA, Amsterdam.
This research was supported by the
Deutsche Forschungsgesellschaft (DFG), the ``Stichting voor
Fun\-da\-men\-teel On\-der\-zoek der Materie (FOM)''
and by the ``Stichting Nationale Computer Faciliteiten (NCF)''.

\end{document}